\documentclass[a4paper,12pt]{article}
\usepackage[english]{babel}
\usepackage{amsmath}%
\usepackage{cite}%
\newtheorem{theorem}{Theorem}
\usepackage{hyperref}
\usepackage{lamsarrow}
\usepackage{pb-diagram,pb-lams} 
\def\eqa#1{\begin{equation}\begin{aligned}#1\end{aligned}\end{equation}}
\def\eq#1{\begin{equation}#1\end{equation}}
\def\eqs#1{\begin{eqnarray}#1\end{eqnarray}}
\def\seq#1{\begin{equation*}#1\end{equation*}}
\def\seqs#1{\begin{equation*}\begin{split}#1\end{split}\end{equation*}}

\def\teqref#1{(T\ref{#1})}

\def\eeqref#1{(E\ref{#1})}

\def\Z {\hbox{\Sets Z}}
\def\dfrac#1#2{\frac{\partial #1}{\partial #2}}
\newtheorem{lemma}{Lemma}

\newcounter{transf}
\newenvironment{transform}[1]{\refstepcounter{transf}$$ #1\eqno{(\hbox{T}\thetransf)}$$} 

\newcounter{meq}
\newenvironment{mequ}[1]{\refstepcounter{meq}$$ #1\eqno{(\hbox{E}\themeq)}$$}

\font\Sets=msbm10

\title{\bf Classification of five-point differential-difference equations II}
\author{{\bf R.N. Garifullin$^{1}$, R.I. Yamilov$^1$ and D. Levi$^2$}
\\$^1$ Institute of Mathematics, Ufa Scientific Center,\\ Russian Academy of Sciences,\\ 112 Chernyshevsky Street, Ufa 450008, Russian Federation
\\$^2$Department of Mathematics and Physics, Roma Tre University \\
and Sezione INFN  {\it Roma Tre},\\
Via della Vasca Navale 84, 00146 Rome, Italy\\
{\sl E-mails: rustem@matem.anrb.ru, RvlYamilov@matem.anrb.ru,}
\\ {\sl decio.levi@roma3.infn.it}}

\begin{document}
\maketitle
\abstract{Using the generalized symmetry method we finish a classification, started in the article [R.N.~Garifullin, R.I.~Yamilov and D.~Levi, Classification of five-point
differential-difference equations, J. Phys. A: Math. Theor. {\bf 50} (2017) 125201 (27pp)],  of integrable  autonomous five-point differential-difference equations. 
The resulting list, up to autonomous point transformations,  contains 14 equations some of which seem to be new. We have found non-autonomous or non-point transformations relating most of the obtained equations among themselves as well as their  generalized symmetries.}

\section{ Introduction}
 
Here we conclude a generalized symmetry classification started in the article \cite{gyl17}.
The generalized symmetry method uses the existence of generalized symmetries as an integrability criterion and allows one to classify integrable equations of a certain class. Using this method some important classes of partial differential equations \cite{msy87,mss91}, of Differential-Difference Equations (D$\Delta$Es)
 \cite{asy00,y06}, and of Partial Difference Equations (P$\Delta$Es) \cite{ly11,gy12} have been classified. 

Other integrability criteria have been introduced to classify integrable P$\Delta$Es, see e.g. the consistency around the cube technique introduced in \cite{nnn,bs02,n02}, whose   results  are presented, for example,   in 
 \cite{ABS,abs2,boll,boll12}. 
 
 A class of P$\Delta$Es, particularly important among recently studied, is given by those equations which are defined on a square, i.e. which relate four neighboring points in the two-dimensional plane. The complete classification of the integrable P$\Delta$Es defined on a square is very difficult to perform.
 
Almost all integrable known P$\Delta$Es have the lowest order associated generalized symmetries given by  integrable evolutionary D$\Delta$Es which are defined on  three-point lattices \cite{x09,ly11,lpsy,gsl,gslqv} and  belong  to the classification presented in \cite{y06,ly97}. This is the classification of Volterra type equations 
\eq{\label{y1} \dot u_n=\Phi(u_{n+1}, u_n, u_{n-1})}
presented in  \cite{y83}, and the resulting list of equations is quite big, see the details in the review article \cite{y06}. Here $\dot u_n$ is the derivative of $u_n$ with respect to a continuous variable $t$ and $n$ is discrete integer variable.

Recently one has obtained examples of P$\Delta$Es defined on the square which have the lowest order generalized symmetries defined on more than three-point lattices \cite{a11,gy12,shl,mx13}.  So an alternative classification that seems  easier to perform is that of integrable five-point D$\Delta$Es
\eq{\label{y0}  \dot u_n=\Psi(u_{n+2},u_{n+1}, u_n, u_{n-1}, u_{n-2}).}
Few results in this line of research are already known, see e.g. \cite{a14,a16,a16_1,gyl16,gyl17}. The integrable P$\Delta$Es are then obtained as B\"acklund transformations of these D$\Delta$Es  \cite{lb,l81,gy15,ly09,gyl17}.
Their construction scheme is discussed in more detail in \cite{gy15} and \cite[Appendix B]{gyl17}. 
The best known  integrable example in this class is the Ito-Narita-Bogoyavlensky (INB) equation \cite{i75,na82,bo88}: 
\eq{\label{INB}\dot u_n=u_n(u_{n+2}+u_{n+1}-u_{n-1}-u_{n-2}).}

 The classification of  five-point lattice equations of the form \eqref{y0} will contain equations coming  from the classification of Volterra type equations \eqref{y1} in two ways. 
On one hand, they appear as equations of the form 
\eq{\label{y2} \dot u_n=\Phi(u_{n+2}, u_n, u_{n-2}).}
If $u_n$ is a solution of \eqref{y2}, then the functions $\tilde u_k= u_{2k}$ and $\hat u_k=u_{2k+1}$ satisfy \eqref{y1} with $k$ instead of $n$. Eq. \eqref{y2} is in fact a three-point lattice equation equivalent to \eqref{y1}.
On other hand they appear as  generalized symmetries of \eqref{y1}. Any integrable Volterra type equation has a five-point  symmetry of the form \eqref{y0}. See the explicit results for Volterra type equations presented for example in \cite{y06,hlrw,sy88,sy90}. 

To avoid those two cases, which are included in  the classification of Volterra type equations, and to simplify the problem, we limit ourselves to consideration here of just equations of the form
\eqa
{ \dot u_n = A(u_{n+1}, u_n,& u_{n-1}) u_{n+2}+B(u_{n+1}, u_n, u_{n-1}) u_{n-2}\\&+ C(u_{n+1}, u_n, u_{n-1}),\label{e1}}
where $A$, $B$ and $C$ are $n$-independent functions of their arguments.
The majority of the examples  of D$\Delta$Es of the form \eqref{y0} known up to now   belong to the class \eqref{e1} \cite{a16_1,th96,mx13,gy12,bo88,b91,i75,na82,ap14,gyl17}. So the class \eqref{e1} is not void.

However also few equations of the Volterra classification \eqref{y1} are included in the five-point classification of the equations of the class \eqref{e1}. They are those polynomial equations which are 
linearly dependent on $u_{n+1}$ and $u_{n-1}$ \cite{y06}. Such equations, rewritten in the form \eqref{y2}, belong to the class \eqref{e1}. Also their five-point symmetries are of the form \eqref{e1}. 

The theory of the generalized symmetry method is well-developed in case of Volterra type equations \cite{y06} and it has been modified for the case of D$\Delta$Es depending on 5 and more lattice points in \cite{a14,a16}. The classification problem of  class \eqref{e1} equations seems to be technically quite complicate. For this reason we use a simpler version of the method, compared with the one presented in \cite{a14,a16}, which has been developed in \cite{gyl17}.
For equations analogous to  \eqref{INB}, which are the first members of their hierarchies, the simplest generalized symmetry has the form \cite{ztof91,mx13,a11,gy12,gyl17}:
\eq{u_{n,\tau}=G(u_{n+4},u_{n+3},u_{n+2},u_{n+1},u_{n},u_{n-1},u_{n-2},u_{n-3},u_{n-4}),\label{gen_sym}}
where $u_{n,\tau}$ denotes $\tau$-derivative of $u_n$. We will use the existence of such symmetry as an integrability criterion.

The problem naturally splits into two cases depending on the form of the functions $A$ and $B$ of \eqref{e1}, see an explanation in Section \ref{sec_th}. 
In \cite{gyl17} we studied the case when the functions $A$ and $B$ in \eqref{e1} satisfied the conditions:
\begin{equation}\label{e2}
A \ne \alpha(u_{n+1},u_n) \alpha(u_n,u_{n-1}) , \quad B \ne\beta(u_{n+1},u_n) \beta(u_n,u_{n-1}) 
\end{equation} for any functions $\alpha$ and $\beta$ of their arguments.
This class is called the {\it Class~I}, and it includes such well-known examples as the
INB equation \eqref{INB} and the discrete Sawada-Kotera equation \cite{th96}. 
In this case the following simple criterion for checking conditions \eqref{e2} takes place:
\begin{equation}\label{e3}
\frac{\partial}{\partial u_n} \frac{a_{n+1} a_{n-1}}{a_n} \neq 0 , \qquad 
\frac{\partial}{\partial u_n} \frac{b_{n+1} b_{n-1}}{b_n} \neq 0 ,
\end{equation}where 
\seq{a_n=A(u_{n+1},u_n,u_{n-1}),\quad b_n=B(u_{n+1},u_n,u_{n-1}), 
 }  see \cite{gyl17}. 

In \cite{gyl17} it was presented, as a result of the classification of the Class I equations, a novel equation:
\eq{{\dot u_n = (u_n^2+1)\left(u_{n+2}\sqrt{u_{n+1}^2+1}-u_{n-2}\sqrt{u_{n-1}^2+1}\right). \label{sroot}}
}
In \cite{gy} it is shown that  in the continuous limit \eqref{sroot} goes into the well-known Kaup-Kupershmidt equation, and its integrability has  been proved by constructing an L--A pair and conservation laws of sufficiently high order. 

In this paper we consider the case when 
\begin{equation}\label{en2}
A = \alpha(u_{n+1},u_n) \alpha(u_n,u_{n-1})\  \hbox{  or  } \ B =\beta(u_{n+1},u_n) \beta(u_n,u_{n-1})
\end{equation} 
for some functions $\alpha$ and $\beta$ of their arguments, i.e. when
\begin{equation}\label{e4}
\frac{\partial}{\partial u_n} \frac{a_{n+1} a_{n-1}}{a_n}= 0 \qquad \mbox{or} \qquad 
\frac{\partial}{\partial u_n} \frac{b_{n+1} b_{n-1}}{b_n}= 0.
\end{equation}
We will call this case {\it Class II}.
A known representative of Class II is given in \cite{gy12}:
\eq{\dot u_n=(u_{n+1}u_n-1)(u_nu_{n-1}-1)(u_{n+2}-u_{n-2})\label{our0}.} 

In this article we present a complete list of equations of the Class~II possessing a generalized symmetry of the form \eqref{gen_sym}. In this way we  complete the classification of integrable equations \eqref{e1} started in \cite{gyl17}. Among them there are a few probably new integrable examples. Then we find the  non-point autonomous or point non-autonomous transformations relating  most of resulting equations among themselves.

In Section \ref{sec_th} we discuss a theory of the generalized symmetry method suitable to solve our specific problem for Class II equations. In particular, in Section \ref{s_class} some integrability conditions are derived  and criteria for checking those conditions are proved. In Section \ref{sec_list} we present the   list of integrable equations and the relations between those equations expressed as autonomous non-point or non-autonomous point transformations. In Section \ref{sec_sym} the generalized symmetries of the resulting equations are discussed. Section \ref{s_con} is devoted to some concluding remarks.

\section{Theory}\label{sec_th}
Here we briefly repeat the theory, presented in the previous paper \cite{gyl17} and necessary for our present work, as well as derive few additional results related just to the Class II case.

To simplify the notation let us represent  \eqref{e1} as:
\eq{\dot u_n=a_nu_{n+2}+b_nu_{n-2}+c_n\equiv f_n,\label{ur_abc}} where \eqa{ a_n=A(u_{n+1},&u_n,u_{n-1}),\quad b_n=B(u_{n+1},u_n,u_{n-1}),\\ &c_n=C(u_{n+1},u_n,u_{n-1}).\label{ur_abc0}}
In \eqref{ur_abc} we require
 \eq{a_n\neq0,\quad b_n\neq0.\label{ogr_AB}} 
  For convenience we denote the symmetry  \eqref{gen_sym} as: 
\eq{ u_{n,\tau}=g_n,\label{10s}}
with the restriction:
\eq{ \dfrac{g_n}{u_{n+4}} \ne 0, \qquad \dfrac{g_n}{u_{n-4}} \ne 0. \label{ogr_G}}

The compatibility condition for \eqref{ur_abc} and \eqref{10s} is
\eqa{&u_{n,\tau,t}-u_{n,t,\tau}\equiv D_t g_{n}-D_{\tau}f_{n}=0.\label{con_cond}}
Here $D_t$ and $D_{\tau}$ are the operators of total differentiation with respect to $t$ and $\tau$  given respectively by:
\eq{D_t=\sum_{k\in\Z} f_k\dfrac{}{u_k}, \quad D_{\tau}=\sum_{k\in\Z} g_k\dfrac{}{u_k}.\label{dt}}
As  \eqref{ur_abc} and \eqref{10s} as well as  the compatibility condition \eqref{con_cond} are autonomous, their form do not explicitly depend on the point $n$. For this reason, we write down for short below the equations and the compatibility condition \eqref{con_cond} at the point $n=0$: $\dot u_0=f_0, \ u_{0,\tau}=g_0,$
\eqs{ D_t g_{0}=D_{\tau}f_0.\label{con_sh}} 

We assume as independent variables the  functions \eq{u_0,u_1,u_{-1},u_2,u_{-2},u_{3},u_{-3}.\ldots\label{i_var}}  The condition \eqref{con_sh} must be identically satisfied for all values of the independent variables \eqref{i_var}. Eq. \eqref{con_sh} depends on the variables $u_{-6},u_{-5},\ldots,u_5,u_6$ and it is an overdetermined system of equations for the unknown function $g_0$, for any given $f_0$.
Using a standard technique of the generalized symmetry method \cite{y06}, we can calculate $g_0$ step by step, obtaining some necessary conditions for the function $f_0$. 

\subsection{General case}
The first steps for the calculation of $g_0$ can be carried out with no  restriction on the form of  $f_0$.

In fact, differentiating \eqref{con_sh} with respect to $u_6$, we obtain as before (see \cite{gyl17})
up to a $\tau$-scaling in \eqref{10s}:
\eq{\label{gp4}\dfrac{g_0}{u_4}=a_0a_2.}
By differentiating \eqref{con_sh} with respect to $u_5$ and taking into account \eqref{gp4}, we 
can define  
\eq{h^+_0 = \dfrac{g_0}{u_3} - a_1\dfrac{f_0}{u_1}-a_0\dfrac{f_2}{u_3},\label{o_h}}
and we can state the following Lemma \cite{gyl17}:
\begin{lemma}If $h^+_0\ne0$, then there exists $\hat\alpha_n=\alpha(u_n,u_{n-1})$, such that ${a_0=\hat\alpha_1\hat\alpha_0}$.\label{th_h}
\end{lemma}

As a consequence of Lemma \ref{th_h} there are two possibilities:
\begin{itemize}
\item Case 1. Let $a_0\neq\hat\alpha_1\hat\alpha_0$ for any $\hat\alpha_n=\alpha(u_n,u_{n-1})$, cf. \eqref{e2}. Then $h^+_0=0$ due to Lemma \ref{th_h}.
\item Case 2. Let $a_0=\hat\alpha_0\hat\alpha_1$ for some $\hat\alpha_n=\alpha(u_n,u_{n-1})$. Then we can find: \eq{\label{hp}h^+_0=\mu^+\hat\alpha_0\hat\alpha_1\hat\alpha_2} with an arbitrary constant $\mu^+$.
\end{itemize}
\noindent In both cases \eqref{o_h} gives us $\dfrac{g_0}{u_3}$.

In quite similar way, differentiating $\eqref{con_sh}$ with respect to $u_{-6}$ and $u_{-5}$, we get a set of relations analogous to \eqref{gp4} and \eqref{o_h}. Namely,
\eq{\label{gm4}\dfrac{g_0}{u_{-4}}=\nu b_0b_{-2},}
where $\nu$ is an arbitrary nonzero constant, and
\eq{h^-_0 = \dfrac{g_0}{u_{-3}} - \nu b_{-1}\dfrac{f_0}{u_{-1}}-\nu b_0\dfrac{f_{-2}}{u_{-3}}.\label{o_hh}}

We can state a Lemma similar to Lemma \ref{th_h} and, as a consequence, we get again  two cases:
\begin{enumerate}
\item Let $b_0\neq\hat\beta_0\hat\beta_{-1}$ for any $\hat\beta_n=\beta(u_{n+1},u_{n})$, then $h^-_0=0$.
\item Let  $b_0=\hat\beta_0\hat\beta_{-1}$, then we can find: \eq{\label{hm}h^-_0=\mu^-\hat\beta_0\hat\beta_{-1}\hat\beta_{-2}} with an arbitrary constant $\mu^-$.
\end{enumerate}
\noindent In both cases \eqref{o_hh} provides us $\dfrac{g_0}{u_{-3}}$.

So the results presented in this subsection provide a natural frame for splitting further calculation into several different cases. Obviously,  any of the equations \eqref{e1} belongs  either to Class~I given by \eqref{e2} or to Class~II given by \eqref{en2}. In the following  in this paper we consider the Class~II defined in the Introduction by \eqref{en2}.

\subsection{Integrability conditions for equations of Class~II}\label{s_class}
We can construct two types of integrability conditions for equations of Class II. The first of them is obtained  when one of the conditions \eqref{en2} is not satisfied. Up to the involution $n\to -n$, this corresponds to the case 
\eq{A\ne \alpha(u_{n+1},u_n) \alpha(u_n,u_{n-1}),\quad  \quad B=\beta(u_{n+1},u_n) \beta(u_n,u_{n-1}),\label{c1} } i.e. $$h^+_0=0,\qquad   h^-_0=\mu^-\hat\beta_0\hat\beta_{-1}\hat\beta_{-2}.$$ 
As we know, in this case the partial derivatives $\dfrac{g_0}{u_3}$, $\dfrac{g_0}{u_{-3}}$, $\dfrac{g_0}{u_4}$ and $\dfrac{g_0}{u_{-4}} $ are given by (\ref{gp4}, \ref{o_h}, \ref{gm4}, \ref{o_hh}).

Differentiating \eqref{con_sh} with respect to $u_4$ and $u_{-4}$ and introducing the functions:
\eq{q^+_0=\frac1{a_0}\dfrac{g_0}{u_2}-D_t \log a_0-\dfrac{f_0}{u_0}-\dfrac{f_2}{u_2}-\frac1{a_0}\dfrac{f_0}{u_1}\dfrac{f_1}{u_2},\label{qp}}
\eq{q^-_0=\frac1{\nu b_0}\dfrac{g_0}{u_{-2}}-D_t \log b_0-\dfrac{f_0}{u_0}-\dfrac{f_{-2}}{u_{-2}}-\frac1{b_0}\dfrac{f_0}{u_{-1}}\dfrac{f_{-1}}{u_{-2}},\label{qm}}
 we obtain two relations. The first of them has the form of a conservation law \cite{gyl17}:
\eq{2D_t \log a_0=q^+_2-q^+_0.\label{la}} 
The second one is more complicated:
\eq{2D_t \log b_0+(T^{-3}-1)\left(\frac{\mu^-}{\nu\hat\beta_0}\dfrac{f_0}{u_{-1}}\right)=q^-_{-2}-q^-_0.\label{lb}}
Relations \eqref{la} and \eqref{lb} provide necessary conditions for the integrability. If  \eqref{ur_abc} is integrable in the sense that a symmetry \eqref{10s} exists, then there must exist the functions $q_n^+,q_n^-$ depending on a finite number of independent variables \eqref{i_var}, such that the relations (\ref{la}, \ref{lb}) are satisfied. 


When both  conditions \eqref{en2} are satisfied, i.e.
\eq{A=\alpha(u_{n+1},u_n) \alpha(u_n,u_{n-1}),\quad  \quad B=\beta(u_{n+1},u_n) \beta(u_n,u_{n-1}), \label{c2}}
then differentiating \eqref{con_sh} with respect to $u_4$ we obtain instead of \eqref{la}  the integrability condition 
 \eq{2D_t \log a_0+(T^{3}-1)\left(\frac{\mu^+}{\hat\alpha_{0}}\dfrac{f_0}{u_{1}}\right)=q^+_2-q^+_0,\label{lla}}     which  is similar to \eqref{lb}.

The relation \eqref{la} has the form of conservation law.
Due to \eqref{c2}, i.e $a_0=\hat\alpha_1\hat\alpha_0,\ b_0=\hat\beta_0\hat\beta_{-1}$, 
 (\ref{lb}, \ref{lla}) can also be represented as conservation laws:
\eq{D_t\log \hat\alpha_0=(T-1)Q^+_0,\qquad D_t\log \hat\beta_0=(T^{-1}-1)Q^-_0,\label{claw}}
where
\seqs{Q_0^+=&\frac 14(T+1)q_0^+-\frac14(T^2+T+1)\left(\frac{\mu^+}{\hat\alpha_{0}}\dfrac{f_0}{u_{1}}\right)-\frac 12 D_t\log\hat\alpha_0,\\
Q_0^-=&\frac 14(T^{-1}+1)q_0^--\frac14(T^{-2}+T^{-1}+1)\left(\frac{\mu^-}{\nu\hat\beta_{0}}\dfrac{f_0}{u_{-1}}\right)-\frac 12 D_t\log\hat\beta_0.} 
 Eqs.  \eqref{la} and \eqref{claw} are conservation laws of the minimum possible order \cite{y06}.
Consequently any integrable equation under consideration must have two conservation laws of the forms  \eqref{la} or \eqref{claw}.

If, for a given equation \eqref{ur_abc},  conditions (\ref{la}, \ref{lb}) in the  case \eqref{c1} or (\ref{lb}, \ref{lla}) in the  case \eqref{c2} are satisfied and the functions $q_0^\pm$ and $Q_0^{\pm}$ are known, then partial derivatives $\dfrac{g_0}{u_2},\dfrac{g_0}{u_{-2}}$ can be obtained from (\ref{qp}, \ref{qm}). So the right hand side of symmetry \eqref{10s} is defined up to one unknown function of three variables:
\eq{\label{fpsi}\psi(u_{n+1},u_n,u_{n-1}).} This function can be found directly from the compatibility condition \eqref{con_cond}.

In this way we carry out the classification of the equations of Class~II. At first  we use the integrability conditions (\ref{la}, \ref{lb}) or (\ref{lb}, \ref{lla}). Then we define the symmetry up to a function \eqref{fpsi} and then try to derive it implementing the compatibility condition \eqref{con_cond}.

To derive simpler relations to check the integrability conditions (\ref{la}, \ref{lb}, \ref{lla}) we use the variational derivatives considered in \cite{gyl17}.

 For any function
\eq{\label{varphi}\varphi=\varphi(u_{m_1},u_{m_1-1},\ldots,u_{m_2}),\quad m_1\geq m_2,} 
 we define the formal variational derivative through the formula:
\eq{\frac{\delta \varphi}{\delta u_0}=\sum_{k=m_2}^{m_1}T^{-k}\dfrac{\varphi}{u_k},\label{v1}} see e.g. \cite{y06}, and its adjoint version \cite{gyl17}: 
\eq{\frac{\bar\delta \varphi}{\bar\delta u_0}=\sum_{k=m_2}^{m_1}(-1)^kT^{-k}\dfrac{\varphi}{u_k}.\label{v2}} 
Then we can state the following Lemma: 
\begin{lemma}\label{l_c}  The equations $\label{fphi}\frac{\delta \varphi}{\delta u_0}=0$ and $ \frac{\bar\delta \varphi}{\bar\delta u_0}=0$ hold  iff $\varphi$ can be represented in the form \eq{\varphi=\kappa+(T^2-1)\omega,\label{ad_f_d}} where $\kappa$ is a constant, while $\omega$ is a function of a finite number of independent variables \eqref{i_var}.
\end{lemma}

To check if  a given function $\varphi$ is of the form $\varphi=(T^2-1)\omega,$  we have at first to check the conditions of Lemma \ref{fphi}. Then we can represent $\varphi$ in the form \eqref{ad_f_d} and check if $\kappa=0$.

So the criteria for checking   \eqref{la} are of the form:
\eq{\label{ca}\frac{\delta }{\delta u_0}D_t\log a_0=0,\quad\frac{\bar\delta }{\bar\delta u_0}D_t\log a_0=0, }see  \cite{gyl17}. 
In the case of the integrability conditions (\ref{lb}) and (\ref{lla}), we first get \seqs{D_t\log a_0=(T+1)D_t\log \hat\alpha_0,\qquad D_t\log b_0=(T^{-1}+1)D_t\log \hat\beta_0} from
 \eq{\label{ab}a_0=\hat\alpha_1\hat\alpha_0,\qquad b_0=\hat\beta_0\hat\beta_{-1}.}
Then from  (\ref{lb}) and (\ref{lla}) we derive the following criteria for checking  them: 
\eq{\label{cb}\frac{\delta }{\delta u_0}D_t\log \hat \beta_0=0,\quad\mu^-\frac{\bar\delta }{\bar\delta u_0}\left(\frac{1}{\hat\beta_0}\dfrac{f_0}{u_{-1}}\right)=0,}
\eq{\label{ca0}\frac{\delta }{\delta u_0}D_t\log\hat\alpha_0=0,\quad \mu^+\frac{\bar\delta }{\bar\delta u_0}\left(\frac{1}{\hat\alpha_{0}}\dfrac{f_0}{u_{1}}\right)=0.}

\subsection{The  classification}\label{th_res}

Using the method described in the previous subsections, we carry out the classification of integrable equations belonging to the Class~II. 
It is easy to proof  that there are no integrable equations in the asymmetric case \eqref{c1}. So here we discuss only the symmetric case \eqref{c2}.

Let us recall that the partial derivatives $\dfrac{g_0}{u_{\pm 4}}$ and $\dfrac{g_0}{u_{\pm 3}} $ are given by (\ref{gp4},   \ref{o_h}, \ref{gm4}, \ref{o_hh}), and that 
we have implicit definitions for $\dfrac{g_0}{u_{\pm 2}}$ given by the relations (\ref{qp}, \ref{qm}, \ref{lb}, \ref{lla}). 

Applying $\dfrac{}{u_{3}}$ and $\dfrac{}{u_{-3}}$ to (\ref{lb}) and (\ref{lla}) respectively, we get:
\eqs{(\nu+1)\dfrac{\hat\alpha_0}{u_{-1}}=0,\qquad(\nu+1)\dfrac{\hat\beta_0}{u_1}=0,\label{c_nu}}
where $\hat\alpha_0,\,\hat\beta_0$ are defined in (\ref{ab}), while the parameter $\nu$ has been introduced in  \eqref{gm4}. By using  \eqref{c_nu} we find that if $\nu\neq -1$, i.e. when  $\dfrac{\hat\alpha_0}{u_{-1}}$ and $\dfrac{\hat\beta_0}{u_1}$ are zero, there are no integrable equations.

In the symmetric case with $\nu=-1$, from (\ref{lb}, \ref{lla})  we can derive some simple  integrability conditions, see also (\ref{cb}, \ref{ca0}).
Differentiating \eqref{lb} and \eqref{lla}, we can find  the second derivatives of $g_0$, i.e. 
$\dfrac{^2g_0}{u_{2}^2},\ \dfrac{^2g_0}{u_{2}\partial u_{-2}},\ \dfrac{^2g_0}{u_{-2}^2}$. 
Then using the following consequences of the compatibility condition \eqref{con_cond}:
\seq{\dfrac{}{u_2}\left(\frac{1}{a_1}\dfrac{^3}{u_3\partial u_{-2}^2}(D_t g_{0}-D_{\tau}f_{0})\right)=0,}
\seq{\dfrac{}{u_{-2}}\left(\frac{1}{b_{-1}}\dfrac{^3}{u_{-3}\partial u_{2}^2}(D_t g_{0}-D_{\tau}f_{0})\right)=0,} together with formulae for the partial derivatives $g_0$ with respect to $u_2$ and $u_{-2}$, we get the conditions:
\eq{\dfrac{\hat\alpha_0}{u_{-1}}\dfrac{^2\hat\alpha_{-1}}{u_{-2}^2}=0,\qquad \dfrac{\hat\beta_0}{u_{1}}\dfrac{^2\hat\beta_{1}}{u_{2}^2}=0.\label{dhat}} 

From \eqref{dhat} we derive the simple integrability conditions:
\eq{\dfrac{^2\alpha(u_0,u_{-1})}{u_{-1}^2}=0,\qquad \dfrac{^2\beta(u_1,u_0)}{u_{1}^2}=0, \label{dalphabeta}}
where the functions $\alpha, \beta$ are  defined in \eqref{c2}.
In such a way we have reduced the  classification to the calculation of four unknown functions of one variable instead of two functions $\alpha,\beta$ of two variables.

In next section we will get all integrable equations of the form (\ref{e1}, \ref{c2})  with $\alpha,\beta$ satisfying \eqref{dalphabeta}. Since $\nu=-1$, from  (\ref{gp4}, \ref{gm4}) we get that the generalized symmetry \eqref{gen_sym} will have the form:
\eqs{u_{0,\tau}=\hat\alpha_0\hat\alpha_1\hat\alpha_2\hat\alpha_3u_4-\hat\beta_{0}\hat\beta_{-1}\hat\beta_{-2}\hat\beta_{-3}u_{-4}+\hat G(u_3,u_2,u_1,u_0,u_{-1},u_{-2},u_{-3}).\label{g_sym}}

\section{Complete list of integrable equations of Class II}\label{sec_list}

In this Section we present the complete list of integrable equations of Class~II together with the non-point autonomous or point non-autonomous relations between them and together with their generalized symmetries. These equations are referred by the numbers \eeqref{eq1}-\eeqref{SK2}. Some of the obtained equations seem to be new. 

The classification is carried out in two steps: at first one finds all integrable equations of a certain class up to point transformations, then one searches for non-point transformations which link the different resulting equations. In this paper we use autonomous point transformations which, because of the specific form \eqref{e1} of the equations, are linear transformations with constant coefficients:\eq{\label{point_t}\hat u_0=c_1u_0+c_2,\quad  \hat t=c_3t,\qquad c_1c_3\neq0.}
The non-point transformations linking the different resulting equations  are transformations of the form
\eq{\label{transf}\hat u_0=\varphi(u_k,u_{k-1},\ldots,u_m),\ \ k>m,} and their compositions. Some of the resulting equations are related among each other by point non-autonomous transformations.

Eq. \eqref{transf} transforms \eqref{y0}  into
\eq{\label{ty0}   \hat u_{0,t}=\hat \Psi(\hat u_{2},\hat u_{1}, \hat u_0, \hat u_{-1}, \hat u_{-2}).}
For any solution $u_n$ of \eqref{y0}, formula \eqref{transf} provides a solution $\hat u_n$ of \eqref{ty0}. 

The  transformation \eqref{transf} is explicit in one direction. If an equation $A$ is transformed into $B$ by a transformation \eqref{transf}, then this transformation has the {\it direction} from $A$ to $B$, and we will write in diagrams below $A\longrightarrow B$, so indicating  the direction in which it is explicit.
Non-autonomous point transformations 
\eq{\label{point_n_t}\hat u_n=\xi_n u_n,\quad  \hat t=ct,\qquad c\neq0,\quad\xi_n \neq0,\ \forall n ,} which are invertible, will be denoted by $\begin{diagram}\node{A}\arrow{e,l,..}{}\node{B}\arrow{w,l,..}{}\end{diagram}$.

The classification result is formulated in the following theorem:

\begin{theorem} If a nonlinear equation of the form (\ref{ur_abc}--\ref{ogr_AB}) belongs to Class~II \eqref{en2} and has a generalized symmetry (\ref{gen_sym}, \ref{10s}, \ref{ogr_G}), then up to an autonomous point transformation \eqref{point_t} it is equivalent to one of the following equations \eeqref{eq1}--\eeqref{SK2}.  Any  equation in \eeqref{eq1}--\eeqref{SK2} has a generalized symmetry of the form (\ref{gen_sym}, \ref{10s}, \ref{ogr_G}).
\end{theorem}

For a better understanding of the results, we split the complete list into the Lists 1-4, where the equations are related among themselves either by autonomous non-point transformations or by non-autonomous point ones. For each of these lists we show the  relations between the equations by a diagram, where the transformations  \eqref{transf} or \eqref{point_n_t} are shown by arrows and are denoted by the numbers \teqref{traz}--\teqref{tVol2}. 

All necessary transformations are given in the List T. Autonomous non-point transformations, which are linearizable \cite{gyl16}, were constructed by using the transformation theory presented in \cite{gyl16}. A shorter version of this theory with some modifications can be found in \cite[Appendix A]{gyl17}.

The generalized symmetries of \eeqref{eq1}-\eeqref{SK2}  are discussed in Section \ref{sec_sym}.

\bigskip\bigskip
\noindent{\bf List 1.} {\it Equations related to the double Volterra equation}

\begin{mequ}{\dot u_0=u_{{0}} \left[u_1(u_2-u_0)+u_{-1}(u_0-u_{-2}) \right]\label{eq1} }\end{mequ}
\begin{mequ}{\dot u_0=u_{{1}}{u_{{0}}}^{2}u_{-1} \left( u_{{2}}-u_{{-2}} \right)\label{eq2} }\end{mequ}
\bigskip

The equation \eeqref{eq2} has been presented in \cite{ap08}.
Both equations of List 1 are transformed into the equation
\eq{\dot u_0=u_{{0}} \left( u_{{2}}-u_{{-2}} \label{Vsh}\right)} as  shown in Diagram \ref{dVol}. 

\eq{\begin{diagram} \label{dVol} 
\node{\eeqref{eq2}}
\arrow{e,l}{\teqref{tpr2}}
\node[1]{\eeqref{eq1}}
\arrow{e,l}{\teqref{tpr}}
\node[1]{\eqref{Vsh}}
\end{diagram}} 
The non-invertible transformations \teqref{tpr} and \teqref{tpr2} are given in the List T below. Transformations \teqref{tpr} and \teqref{tpr2} are of the  
linearizable class, i.e. are not of Miura type (see a discussion of this notion in \cite{gyl16,gyl17}). 
Eq. \eqref{Vsh} is called the double Volterra equation, as two transformations $\tilde u_k=u_{2k}$ and $\hat u_k=u_{2k+1}$ turn it into the standard form of the  Volterra equation \cite{y06}. 

\bigskip\bigskip
\noindent{\bf List 2.} {\it Equations related to a generalized symmetry of the  Volterra equation}

\begin{mequ}{\dot u_0=u_{{0}} \left[u_1(u_2+u_1+u_0)-u_{-1}(u_0+u_{-1}+u_{-2}) \right]+cu_{{0}} \left( u_{{1}}-u_{{-1}} \right)  \label{Vol1} }\end{mequ}
\begin{mequ}{\dot u_0=(u_{{0}}^2-a^2) \left[(u_1^2-a^2)(u_2+u_0)-(u_{-1}^2-a^2)(u_0+u_{-2}) \right]+c(u_{{0}}^2-a^2) \left( u_{{1}}-u_{{-1}} \right)  \label{mVol2} }\end{mequ}
\begin{mequ}{\dot u_0=(u_1-u_0+a)(u_0-u_{-1}+a)(u_2-u_{-2}+4a+c)+b\label{Volz} }\end{mequ}
\begin{mequ}{\dot u_0=u_0[u_1(u_2-u_1+u_0)-u_{-1}(u_0-u_{-1}+u_{-2})]\label{Vol_mod1} }\end{mequ}
\begin{mequ}{\dot u_0=(u_{{0}}^2-a^2) \left[(u_1^2-a^2)(u_2-u_0)+(u_{-1}^2-a^2)(u_0-u_{-2}) \right]  \label{mVol3} }\end{mequ}
\begin{mequ}{\dot u_0=(u_1+u_0)(u_0+u_{-1})(u_2-u_{-2})\label{Vol_mod2} }\end{mequ}
\bigskip

The equations of the List 2 are related among themselves by the transformations shown in the following diagrams \ref{dgenV}.

\eqs{\label{dgenV}\begin{diagram} \label{d2} \node{\eeqref{Volz}}
\arrow{e,l}{\teqref{traz}}
\node[1]{\eeqref{Vol1}}
\node[1]{\eeqref{mVol2}}\arrow{w,l}{\teqref{tVol}}\\ 
\node[1]{\eeqref{Vol_mod1}}
\arrow{e,l,..}{\teqref{m_ed}}
\node[1]{(E\ref{Vol1},\ c=0)}\arrow{w,l,..}{}\\
\node[1]{\eeqref{mVol3}}
\arrow{e,l,..}{\teqref{tVol2}}
\node[1]{(E\ref{mVol2},\ c=0)}\arrow{w,l,..}{}\\
\node[1]{\eeqref{Vol_mod2}}
\arrow{e,l,..}{\teqref{m_ed}}
\node[1]{(E\ref{Volz},\ a=b=c=0)}\arrow{w,l,..}{}
\end{diagram}} 
Equations \eeqref{Vol1}-\eeqref{Volz} are the generalized symmetries of the  equations \cite{y06}:
\eqs{\dot u_0&=&u_0(u_1-u_{-1})\label{V0},\\ \dot u_0&=&(u_0^2-a^2)(u_1-u_{-1})\label{V1},\\ \dot u_0&=&(u_1-u_0+a)(u_0-u_{-1}+a)\label{V2}.} 
These Volterra type equations  are related among themselves by the same transformations \teqref{traz} and \teqref{tVol} as their symmetries. The transformations \teqref{traz} and \teqref{tVol} are well-known, see e.g. \cite{y06}. Eq. \eqref{V0} is the Volterra equation itself,  \eqref{V1} is the modified Volterra equation, and the transformation \teqref{tVol} with $a\neq0$ is of Miura type \cite{gyl16}. The transformation \teqref{traz} is linear. 

Transformations \teqref{m_ed} and \teqref{tVol2} are non-autonomous invertible point transformations. So \eeqref{Vol_mod1}-\eeqref{Vol_mod2} are equivalent to \eeqref{Vol1}-\eeqref{Volz}, as shown in diagrams \ref{dgenV}. The transformation \teqref{tVol2} is nontrivial, see a comment after List~T. It can be shown that  \eeqref{Vol_mod1}-\eeqref{Vol_mod2} are also generalized symmetries of some simpler non-autonomous equations of the Volterra type.

\bigskip\bigskip
\noindent{\bf List 3.} {\it Equations related to the  INB equation \eqref{INB}}

\begin{mequ}{\dot u_0=u_0(u_2u_1-u_{-1}u_{-2})\label{INB1} }\end{mequ}
\begin{mequ}{\dot u_0=(u_1-u_0+a)(u_0-u_{-1}+a)(u_2-u_1+u_{-1}-u_{-2}+2a)+b\label{INB3} }\end{mequ}
\begin{mequ}{\dot u_0=u_0(u_1u_0-a)(u_0u_{-1}-a)(u_2u_1-u_{-1}u_{-2})\label{MX2} }\end{mequ}
\begin{mequ}{\dot u_0=(u_1+u_0)(u_0+u_{-1})(u_2+u_1-u_{-1}-u_{-2})\label{INB2} }\end{mequ}
\bigskip

Eq. \eeqref{INB1} is a well-known modification of  INB  \eqref{INB}, see \cite{bo88}. Eq. \eeqref{MX2} with $a=0$ has been considered in \cite{ap08}. The equations of  this list are related among themselves and to
\eq{\dot u_0=(u_0^2+au_0)(u_2u_1-u_{-1}u_{-2})\label{MX}}
as shown in the following diagrams:
\eqa{\label{INBd} \begin{diagram} 
\node{\eeqref{INB3}}
\arrow{e,l}{\teqref{traz}}
\node{\eeqref{INB1}}
\arrow{e,l}{\teqref{tpr}}
\node{\eqref{INB}}\\
\node[2]{\eeqref{MX2}}\arrow{e,l}{\teqref{tpr3}}
\node{\eqref{MX}}\arrow{n,l}{\teqref{tm3}}\end{diagram} \\
\begin{diagram} 
\node{\eeqref{INB2}}\arrow{e,l,..}{\teqref{m_ed}}\node{(E\ref{INB3},a=b=0)}\arrow{w,..}
\end{diagram}} 
Eq. \eqref{MX} is presented in \cite{bo88} in the case $a=0$ and in \cite{ap14,mx13} in the case $a\neq 0$. The transformation \teqref{m_ed} is invertible, while all the other transformations are non-invertible. All the transformations present in the diagram \eqref{INBd} are linearizable except for \teqref{tm3} with $a\neq0$ which is of Miura type, see a comment in \cite{gyl16}.


\bigskip\bigskip
\noindent{\bf List 4.} {\it The remaining equations}

\begin{mequ}{\dot u_0=(u_1u_0-1)(u_0u_{-1}-1)(u_2-u_{-2})\label{our} }\end{mequ}
\begin{mequ}{\dot u_0=u_1u_0^3u_{-1}(u_2u_1-u_{-1}u_{-2})-u_0^2(u_1-u_{-1})\label{SK2} }\end{mequ}
\bigskip
Eq. \eeqref{our} is known (see \cite{gy12,gmy14}), while \eeqref{SK2} is a simple modification of 
\eq{\dot u_0=u_0^2(u_2u_1-u_{-1}u_{-2})-u_0(u_1-u_{-1})\label{SK}} as shown in diagram \ref{dSK}.
\eq{\begin{diagram} \label{dSK} 
\node{\eeqref{SK2}}
\arrow{e,l}{\teqref{tpr}}
\node[1]{\eqref{SK}}\end{diagram}}
Eq. \eqref{SK} has been found in \cite{th96} and can be called  the discrete Sawada-Kotera equation \cite{th96,a11}. 

\bigskip
All transformations relating the equations of Lists 1-4 are presented in the  List T. 

\bigskip\bigskip
\noindent{\bf List T.} {\it List of used transformations}

\begin{transform}{\hat u_0=u_1-u_{0}+a\label{traz}}\end{transform}
\begin{transform}{\hat u_0=u_1u_{0}\label{tpr}}\end{transform}
\begin{transform}{\hat u_0=u_1u_{0}-a\label{tpr3}}\end{transform}
\begin{transform}{\hat u_0=u_1u_{-1}\label{tpr2}}\end{transform}
\begin{transform}{\hat u_0=(u_1-a)(u_0+a)\hbox{ or } \hat u_0=(u_1+a)(u_0-a)\label{tVol}}\end{transform}
\begin{transform}{\hat u_0=(u_2+a)u_1u_{0} \hbox { or }\hat u_0=u_2u_1(u_{0}+a)\label{tm3}}\end{transform}
\begin{transform}{\hat u_n=(-1)^n u_{n},\quad \hat t=-t\label{m_ed}}\end{transform}
\begin{transform}{\hat u_n=\kappa_n u_n,\quad \kappa_n=\frac12(1-i)[i^n+i(-i)^n],\quad \hat t=-t\label{tVol2}}\end{transform}
\bigskip
Here transformation \teqref{traz} is linear, while transformations (T\ref{tpr} -- T\ref{tpr2}) and (T\ref{tVol}, T\ref{tm3}) with $a=0$ are linearizable. Transformations (T\ref{tVol}, T\ref{tm3}) with $a\neq 0$ are of Miura type. 

Transformations (T\ref{m_ed}, T\ref{tVol2}) are invertible and non-autonomous. The function $\kappa_n$ appearing in (T\ref{tVol2}) is four-periodic, i.e. $\kappa_{n+4}=\kappa_n$ for all $n$. It can be defined by the following initial conditions: 
\eq{\kappa_0=\kappa_1=1,\quad\kappa_2=\kappa_3=-1} and satisfy the relations:
\eq{\kappa_{n+2}=-\kappa_n,\quad \kappa_n^2=1.}

\subsection{Generalized Symmetries }\label{sec_sym}

There is no need to write down here the generalized symmetries  \eqref{g_sym} for equations \eeqref{eq1}--\eeqref{SK2}) explicitly, as symmetries of key equations are known, while the  symmetries for the other equations can be constructed by the transformations (T\ref{traz} -- T\ref{tVol2}).

In order to construct generalized symmetries of the form \eqref{g_sym}, we need symmetries of equations (\ref{INB}, \ref{Vsh}, \ref{MX}, \ref{SK}) which are given explicitly in \cite[Section 4]{gyl17}. Originally, the generalized symmetry of \eqref{INB} has been presented in \cite{ztof91}, of \eqref{MX} with $a=1$ in \cite{mx13}, of \eqref{MX} with $a=0$ in \cite{ztof91}, and of \eqref{SK} in \cite{a11}. The generalized symmetry of \eeqref{our} can  be found in \cite{gy12,gmy14}. Symmetries of \eeqref{Vol1} and \eeqref{mVol2} belong to the hierarchies of the Volterra and modified Volterra equations which are well-known, see \cite{y06}. 

The generalized symmetries for  (E\ref{Vol_mod1} --  E\ref{Vol_mod2}, E\ref{INB2}) can be constructed  easily with the help of the  invertible transformations (T\ref{m_ed}, T\ref{tVol2}). These transformations are non-autonomous, but in this case they allow us to construct  autonomous symmetries of the form \eqref{g_sym}.

The generalized symmetries for the remaining equations of Lists 1-4, namely (E\ref{eq1}, E\ref{eq2}, E\ref{Volz}, E\ref{INB1} -- E\ref{MX2}, E\ref{SK2}), can be constructed by using the non-invertible transformations (T\ref{traz}--T\ref{tpr2}), where \teqref{tpr} is a particular case of \teqref{tpr3}.
For the sake of clarity we present here the construction scheme for the generalized symmetries obtained by the transformations (T\ref{traz}, T\ref{tpr3}, T\ref{tpr2}), see more details in \cite{gyl16,gyl17}.

Let us first consider the case when an equation $A$ is transformed into an equation $B$ by transformation \teqref{traz}: $\begin{diagram}  \node{A}
\arrow{e,l}{\teqref{traz}}
\node{B.}\end{diagram}$  If $B$ has a symmetry 
\eq{\hat u_{0,\tau}=\hat G(\hat u_{4},\hat u_{3},\hat u_{2},\ldots,\hat u_{-4}),\label{gen_sym_hat}} 
then
we look for a symmetry of the form \eqref{gen_sym} for  $A$. As $\hat u_{0,\tau}=(T-1)u_{0,\tau},$ we should represent the function $\hat G$ in the form
\eq{\hat G=(T-1)H(\hat u_{3},\hat u_{2},\ldots,\hat u_{-4})}
and then we immediately get \eqref{gen_sym}:
\eq{u_{0,\tau}=G=H|_{\hat u_k=u_{k+1}-u_{k}+a}.}
It is evident that $G$ contains an arbitrary constant of integration, as it is obtained by solving a first order difference equation.

Let us now consider the case: $\begin{diagram}  \node{A}
\arrow{e,l}{\teqref{tpr2}}
\node{B.}\end{diagram}$  As $(\log \hat u_{0})_{\tau} = (T+T^{-1})(\log u_{0})_\tau,$ then for \eqref{gen_sym_hat} we should get the representation 
\eq{{\hat G}/{\hat u_0}= (T+T^{-1})H(\hat u_{3},\hat u_{2},\ldots,\hat u_{-3}),}
and consequently \eqref{gen_sym} is given by:
\eq{u_{0,\tau}=G=u_0 H|_{\hat u_k=u_{k+1}u_{k-1}}.}
No constant of integration arises in this case.

In the case $\begin{diagram}  \node{A}
\arrow{e,l}{\teqref{tpr3}}
\node{B}\end{diagram}$ we have: $(\log (\hat u_{0}+a))_{\tau} = (T+1)(\log u_{0})_\tau.$ Equation \eqref{gen_sym_hat}  turns out to be:  
\eq{{\hat G}/{(\hat u_0+a)}= (T+1)H(\hat u_{3},\hat u_{2},\ldots,\hat u_{-4}),\label{gT3}}
and \eqref{gen_sym} is given by:
\eq{u_{0,\tau}=G=u_0 H|_{\hat u_k=u_{k+1}u_{k}-a}.\label{symT3}}
No constant of integration arises here.

Now we consider the example: $\begin{diagram}  \node{\eeqref{eq1}}
\arrow{e,l}{\teqref{tpr}}
\node{\eqref{Vsh}.}\end{diagram}$ The symmetry for \eqref{Vsh} is:
\seq{\hat u_{0,\tau}=\hat u_0[\hat u_2(\hat u_4+\hat u_2+\hat u_0)-\hat u_{-2}(\hat u_0+\hat u_{-2}+\hat u_{-4})],} see \cite{gyl17}. The transformation \teqref{tpr} is \teqref{tpr3} with $a=0$, and we need to get the representation \eqref{gT3}. As
\seq{\hat G/\hat u_0=(T^2-T^{-2})\hat u_0(\hat u_2+\hat u_0+\hat u_{-2})} and
\seq{T^2-T^{-2}=(T+1)(T-1)(1+T^{-2}),}  the generalized symmetry for \eeqref{eq1} is given by:
\seq{u_{0,\tau}=u_0(T-1)(1+T^{-2})u_1u_0(u_3u_2+u_1u_0+u_{-1}u_{-2}).}

\section{Conclusion} \label{s_con}
Here we have finished a generalized symmetry classification started in our previous article \cite{gyl17}. 
The resulting list contains 14 equations, some of which seem to be new. We have found non-autonomous or non-point transformations
 relating most of the resulting
equations among themselves.

Using the obtained five-point integrable equations \eqref{e1}, we can construct integrable examples of partial difference
equations, defined on a square lattice. Their construction scheme is discussed in \cite{gy15} and \cite[Appendix B]{gyl17}. 

Connections between different equations obtained in this paper are simpler than in the previous article \cite{gyl17}. However, a transformation can relate two five-point equations \eqref{e1}, but this connection may collapse for the corresponding P$\Delta$Es. So it may happen that for one of the five-point equations the corresponding P$\Delta$E exists, while for another  the corresponding P$\Delta$E becomes nonlocal. Therefore the problem of the construction of P$\Delta$Es remains nontrivial.

\end{document}